\documentclass[conference]{IEEEtran}
\IEEEoverridecommandlockouts
\usepackage{amsmath,amssymb,amsfonts}
\usepackage{algorithmic}
\usepackage{graphicx}
\usepackage{textcomp}
\usepackage{xcolor}
\def\BibTeX{{\rm B\kern-.05em{\sc i\kern-.025em b}\kern-.08em
    T\kern-.1667em\lower.7ex\hbox{E}\kern-.125emX}}

\usepackage{biblatex} 
\begin{document}

\title{MAO: Machine learning approach for NUMA optimization in Warehouse Scale Computers}

\author{\IEEEauthorblockN{1\textsuperscript{st} Yueji Liu}
\IEEEauthorblockA{\textit{Baidu Group} \\
Beijing, China \\
liuyueji@baidu.com}
\and
\IEEEauthorblockN{2\textsuperscript{nd} Jun Jin}
\IEEEauthorblockA{\textit{Intel Corporate} \\
Shanghai, China \\
jun.i.jin@intel.com}
\and
\IEEEauthorblockN{3\textsuperscript{rd} Wenhui Shu}
\IEEEauthorblockA{\textit{Intel Corporate} \\
Shanghai, China \\
kevin.shu@intel.com}
\and
\IEEEauthorblockN{4\textsuperscript{th} Shiyong Li}
\IEEEauthorblockA{\textit{Baidu Group} \\
Beijing, China \\
lishiyong@baidu.com}
\and
\IEEEauthorblockN{5\textsuperscript{th} Yongzhan He}
\IEEEauthorblockA{\textit{Baidu Group} \\
Beijing, China \\
heyongzhan@baidu.com}
}

\maketitle

\begin{abstract}
Non-Uniform Memory Access (NUMA) architecture imposes numerous performance challenges to today’s cloud workloads. Due to the complexity and the massive scale of modern warehouse-scale computers (WSCs), a lot of efforts need to be done to improve the memory access locality on the NUMA architecture. In Baidu, we have found that NUMA optimization has significant performance benefit to the top major workloads like Search and Feed(Baidu’s recommendation system). But how to conduct NUMA optimization within the large scale cluster brings a lot of subtle complexities and workload-specific scenario optimizations. In this paper, we will present a production environment deployed solution in Baidu called MAO (Memory Access Optimizer) that helps improve the memory access locality for Baidu’s various workloads. MAO includes an online module and an offline module. The online module is responsible for online monitoring, dynamic NUMA node binding and runtime optimization. Meanwhile the offline workload characterization module will proceed with the data analysis and resource-sensitivity model training. We also proposed a new performance model called "NUMA Sensitivity model" to address the impact of remote memory access to workload performance and projection of the potential performance improvements via NUMA optimization for specified workload. Based on continuous data collected from online monitoring, this model is proved to be working properly in MAO. As of today, we have successfully deployed MAO to more than one hundred thousand servers. In our Feed product, we have achieved 12.1\% average latency improvements and 9.8\% CPU resource saving. 
\end{abstract}

\begin{IEEEkeywords}
Memory Access Optimizer, Warehouse Scale Computers, NUMA Sensitivity model, Machine Learning, XGBoost
\end{IEEEkeywords}

\section{Introduction}
Non-Uniform Memory Access (NUMA) architecture imposes numerous performance challenges to today’s cloud workloads. Due to the complexity and modern warehouse scale computers (WSCs), lots of engineering efforts need to be conducted to improve the memory access locality based on NUMA architecture. In this case, requests issued from a CPU core can either access its own NUMA node’s DRAM (known as local NUMA access) or access other NUMA nodes through cross-socket or cross-die interconnection. Though the interconnection is designed to transfer data at a relative high speed, remote memory access latency is still a non-negligible factor for workload's performance. And this factor might be even significant in a multi-sockets system that there are more than one hops needed. Traditional optimization to NUMA architecture is always focus on improving the memory locality by threads or memory pages migration. 
Achieving optimal performance requires optimal placement of each thread to be close to the data it accesses\cite{antony2006exploring}\cite{brecht1993importance}. However, such migrations have to be conducted after looking at the overall load-balancing among all cores and memory in the machine in order to maintain a good utilization of the hardware resources. Multi-threaded applications may distribute many threads on distant cores of a single host and memory among all memory nodes\cite{goglin2009enabling} as well. 
A tremendous amount of research effort has been devoted to investigate the impact of NUMA and propose optimization algorithms on data placement and OS scheduling. 
\begin{itemize}
    \item There has been a wealth of prior research regarding NUMA related scheduling approach\cite{lepers2014improving}\cite{gaud2015challenges}\cite{psaroudakis2016adaptive}\cite{brecht1993importance}\cite{verma2015large}\cite{majo2011memory}\cite{mccormick2011empirical} \cite{luo2016compositional}. These related work specifically focused on core allocation and data placement strategy for NUMA optimization. \par
For example, AutoNUMA\cite{corbet2012autonuma} improves memory access locality by migrating threads on the nodes they access the most and pages on the node from which they are most accessed based on hot page profiling mechanism. DINO\cite{blagodurov2010case} made some more progress and designed a new contention-aware scheduling algorithm based on LLC MPKI (last level cache misses per 1000 instructions) for NUMA systems to minimize cache contention while avoiding interconnect and memory controller contention. Though DINO eliminated superfluous migrations but it doesn’t address the data sharing issue between threads which is quite common in cloud workloads. Carrefour\cite{lepers2014improving}\cite{gaud2015challenges} collected performance metrics and page accesses statistics and worked out a novelty to combine 3 main strategies including page migrating, replicating and page interleaving. However, as it still highly relies on page migration technology that may incur high overhead especially in WSCs.\par
    \item Besides, Machine learning  approaches\cite{denoyelle2019data}\cite{funston2018placement}\cite{arapidis2018performance} were proposed to predict performance and the sensitivity of applications. However, these approaches either rely on offline binary instrumentation or need extensive experiments to different placement policy of threads and configurations to collection training samples, which is not scalable to thousands of workloads at WSCs.\par
    \item Previous work proposed a light weight NUMA score (sensitivity) model was proposed in \cite{tang2013optimizing} based on the status of CPU utilization and memory usage per NUMA node, which might be stale and inaccurately predicate the performance impact of NUMA locality. In Baidu, we have found that NUMA sensitivity is more correlated with remote memory bandwidth ratio derived by the percentage of remote memory bandwidth for a specific workload to overall memory bandwidth, more details will be discussed in Chapter 3.5. \par
\end{itemize}
When we inspect into our production environment deployment, we found the existing solutions might not fit well. In this paper, we addressed the aforementioned challenges in WSCs, presenting our system design called MAO (Memory Access Optimizer). This work is based on the collaboration with Intel in order to improve the memory access locality for thousands of workloads in production. In summary, we make the following contributions in this paper: \par
\begin{itemize}
    \item We found NUMA binding could be more effective than page migration. In production environment, our study on Search workload shows a lot of online work is getting done by threads that run for 0.4ms to 10ms, which resulted frequent thread migration between different cores. That further makes AutoNUMA less effective. Meanwhile, large remote memory footprint may result in high overhead by page migration. \par
    \item To maximum the utilization the CPU resource, not all service containers can be deployed with NUMA optimization. We proposed an effective NUMA sensitivity model using XGBoost algorithm based on collected workload hardware performance counters. With this model, we observed around 5\% mean absolute error(MAE) to project performance improvements with NUMA optimization and use the info to guide the scheduler based on NUMA awareness or maximum cpu resource. \par
    \item We proposed a dynamic mechanism combines online execution and offline training to address the problem of imbalance CPU usage and resource fragmentation dynamically due to the NUMA binding in order to improve overall system performance.\par
    \item We implemented MAO as a totally solution of NUMA optimization for production deployment. In the past one year, we have successfully deployed MAO to more than 100K servers in Baidu’s WSC optimizing many products like Search and Feed with very promising performance gain, for example, in Feed product, we have achieved 12.1\% average latency improvements and 9.8\% CPU resource saving in average.  
\end{itemize}

\section{Background and Motivation}
\subsection{NUMA in modern CPU Architecture introduction}
This section describes about the NUMA and cross socket interconnection related concepts and terminology in modern x86 micro architecture. We also introduce the fundamental performance difference between UMA and NUMA related optimization. \par
Typically, the CPU socket and the closest memory banks built a NUMA Node. Whenever a CPU needs to access the memory of another NUMA node, it cannot access it directly but is required to access it through the interconnect between the CPU sockets. Figure 1 shows schematic view of a typical two sockets system with multiple CPU/Memory couples equivalent to the socket numbers. Each couple represents a NUMA node, which is the fundamental optimization object of the workload to be aware of regarding data access locality. 
\begin{figure}[h]
    \centering
    \includegraphics[width=0.45\textwidth]{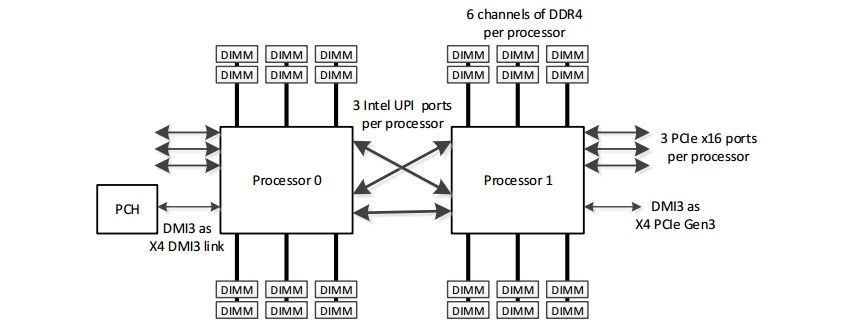}
    \caption{A schematic view of a system with 2 memory domains and 3 UPI supported topology design}
    \label{fig:numa_node_overview}
\end{figure}
In modern CPU micro-architecture, data access from memory will be cached. Depends on what cache level the data is located, there is a clear data access latency difference between each hierarchy layer. from the experiment we did with Intel Memory Latency Checker (MLC) tool\cite{viswanathan2013intel}, the memory latency in nanoseconds on a Xeon SP Skylake Processor with DDR4 2400Mhz memory installed has the access latency to remote memory (138ns) around 72\% higher than access to the local memory (80ns).  

In summary, from the study of Google’s NUMA experience\cite{tang2013optimizing} and our internal experiments based on real workloads, we can tell that NUMA optimization with improved memory access locality can bring significant performance improvements which is the major the target of MAO.

\subsection{Motivation of MAO}
\subsubsection{Warehouse Scale Computers introduction}
Today, we have been supporting thousands of Baidu's products within our warehouse scale computers consists of a hundred thousands of servers. The resources within the fleet have been managed by a cluster operating system called "Matrix" which supports advanced capabilities like job mix deployment, resource isolation, service integration, deployment and failure over, etc. \par
\begin{figure}[h]
    \centering
    \includegraphics[width=0.5\textwidth,height=5cm]{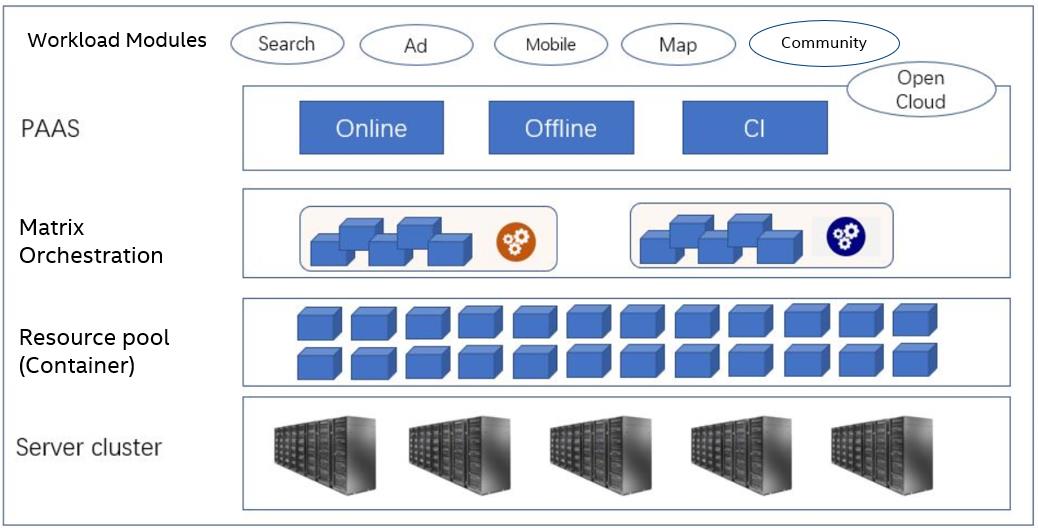}
    \caption{Baidu's Production Warehouse Scale Compute overview}
    \label{fig:CompanyX_production_env_overview}
\end{figure}

As illustrated in Figure 2: In Baidu, we have several key applications called Search, Ad (Advertisement), Mobile, Map, etc. Each application may contain many different services. Take search application for example, it has services including Indexing, Basic Search for online matching, Ranking, Web Tier, etc. Each service is deployed in form of lots instances spread on different servers and each instance can only be deployed in one container which is the smallest resource allocation unit by cluster scheduler.

\subsubsection{Characterization to production environment}
\begin{itemize}
    \item \textbf{Various workloads with different characters.} The SLA target and sensitivity to the NUMA architecture is different from each other. For example, some workloads are compute bound while some are memory bound. Within memory bound workloads, there are still sub categories like memory latency bound and memory bandwidth bound, etc. So how to find a workload SLA agnostic NUMA sensitivity metric in the large scale production environment to properly address the workload preference to NUMA locality or memory bandwidth is a key target we want to address. \\
    After exploration, we decide to leverage an offline modeling approach with collected workloads’ hardware PMU metrics \cite{guide2011intel} to build up a NUMA sensitivity model. With the model, we can find those workloads that are sensitive to the memory locality to conduct the NUMA optimization and calibrate the model with optimization result’s feedback. Eventually, this fine-tuned offline model can also help us to accurately project the NUMA benefits during the runtime.  
    \begin{figure}[h]
        \centering
        \includegraphics[width=0.45\textwidth]{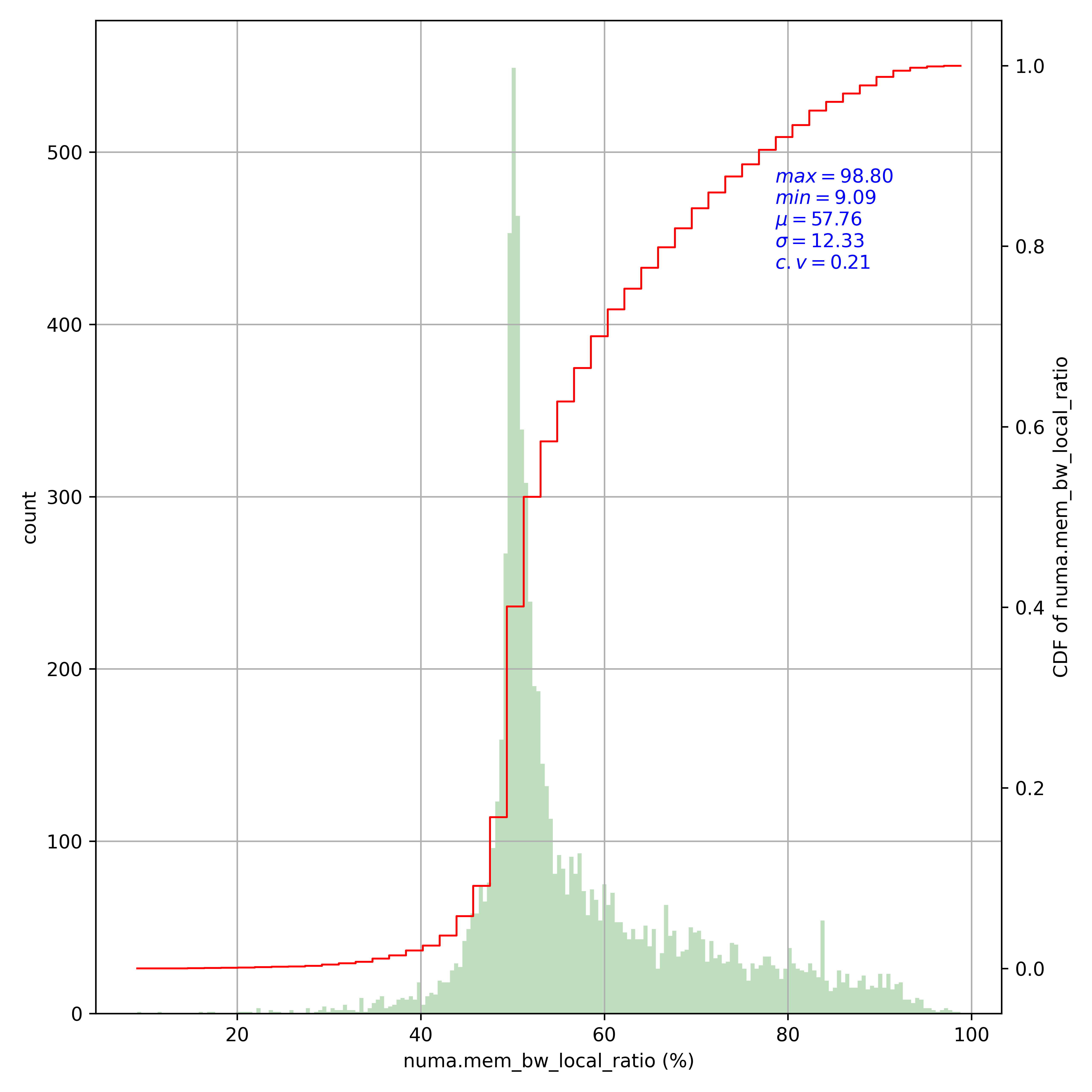}
        \caption{Baidu’s NUMA\_local\_access\_ratio statistic within the cluster with AutoNUMA enabled} 
        \label{fig:NUMA local access ratio statistic}
    \end{figure}

    \item \textbf{Multi-threads workloads with frequently migration during sleep and wake up.} Our kernel scheduler trace result shows that the bulk of the run time between context switches aka the scenario where tasks are switching back and forth are \textless 100us, and a lot of our “work” is getting done by threads that run for 0.4ms to 10ms. So, if we apply memory page migration to support better memory access locality, the migrated page will turn from local access to remote access quickly along with the threads’ migration between the sockets. This negative action will impact the online workload's NUMA local access ratio as shown in Figure3. Due to less effective optimization strategy in memory migration, With AutoNUMA enabled configuration, the average access local ratio only shows a little better than 50\% average to 57.76\% which is much less than our expectation.  
    \begin{figure}[h]
    \centering
    \includegraphics[width=0.45\textwidth]{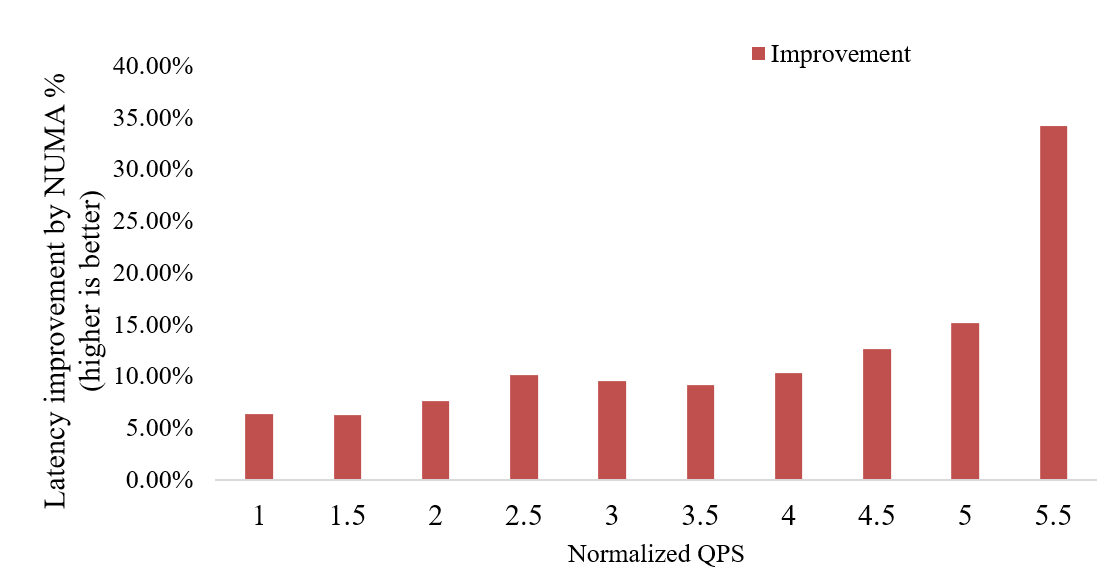}
    \caption{Latency improvements with NUMA optimization for online Search workload}
    \label{fig:numa_node_overview}
    \end{figure}
    
    However, since the CPU resources are shared and dynamically allocated across the workload instances, we found out that with proper NUMA node binding, we can get better memory access locality. And regard to Baidu’s search workload, we confirm the benefit by deploying 4 search instances in the test environment. The further comparison shows in Figure4 tells us that NUMA bind (two instances binded to each NUMA node) has better search performance benefit comparing to autoNUMA in general. The higher normalized stress qps, the more improvements NUMA bind has comparing to AutoNUMA.   
    \item \textbf{Each workload service has multiple instances mix deployed within different servers.} Even with the container isolation, there’s still micro-architecture performance impact against each other. So each instance's placement decision needs to be considered when we conduct NUMA optimization. On one hand, with the trending of resource pooling, the density of mix-deployment gets higher which makes the local memory access more difficult to be guaranteed. On the other hand, we also need to consider the fragmentation and load balance within the server along with NUMA locality optimization. So the way to select candidates from production environment for NUMA optimization is very critical, it can either be based on specific services (one service is deployed in form of multiple instances on different servers) or dynamically select instances deployed on a target server.   \par
    Eventually we decide to conduct NUMA optimization based on service granularity as the solution here. Because usually all of the services require uniformed performance, our solution need to avoid performance variation between different instances of the same service, while instance based selection may result in worse stability and predictability. 
    \item \textbf{The loading of the instances are dynamic, usually periodically by days.} there are peak and valley traffic time in online production environments. Based on our experience, the load imbalance problem may be more prominent on peak traffic time. We need to consider about this along with our NUMA optimization, so the optimization need to be dynamically adjusted. 
    \item \textbf{Stringent overhead criteria of minimize the impact the online workloads.} After the evaluation, we found that the cost of page migration is higher than our expectation that very difficult to be adopted in online production. Below chart describe about the remote memory distribution of one key workload within Baidu that remote memory size range from 51MB to 6GB with average at 2.5GB. Afterwards, We conducted the memory migration experiment based on that shows the efforts to migrate 2GB memory pages, which requires at least one 2.7GHz cpu core to spend 1.3 second at 100\% utilization. The overhead like this is quite unacceptable during the workload’s running. 
    \begin{table}[h!]
    \centering
    \resizebox{8cm}{1.7cm}{
    \begin{tabular}{||c | c ||} 
     \hline
     Migrated memory size & Average Time Spent (second) \\ [0.25ex] 
     \hline\hline
     16MB & 0.007919   \\ 
     32MB & 0.020367   \\
     51MB & 0.033931   \\
     64MB & 0.041946   \\
     128MB & 0.083136    \\
     256MB & 0.162868    \\
     512MB & 0.322639   \\
     1024MB & 0.644564   \\
     2048MB & 1.272319    \\
     2560MB & 1.583247   \\
     \hline
    \end{tabular}
    }
    \caption{Average time taken to migrate memory pages of different total size}
    \label{table:data}
    \end{table}
\end{itemize}

As a result, we are considering to implement a holistic solution across WSCs called “MAO” to improve the current limitation of default NUMA optimization and also interact with existing Matrix cluster orchestration management with sub modules like offline workload analyzer, online monitoring, allocation and optimization, etc.

\section{MAO platform Design and Implementation}

\begin{figure}[h]
    \centering
    \includegraphics[width=0.45\textwidth]{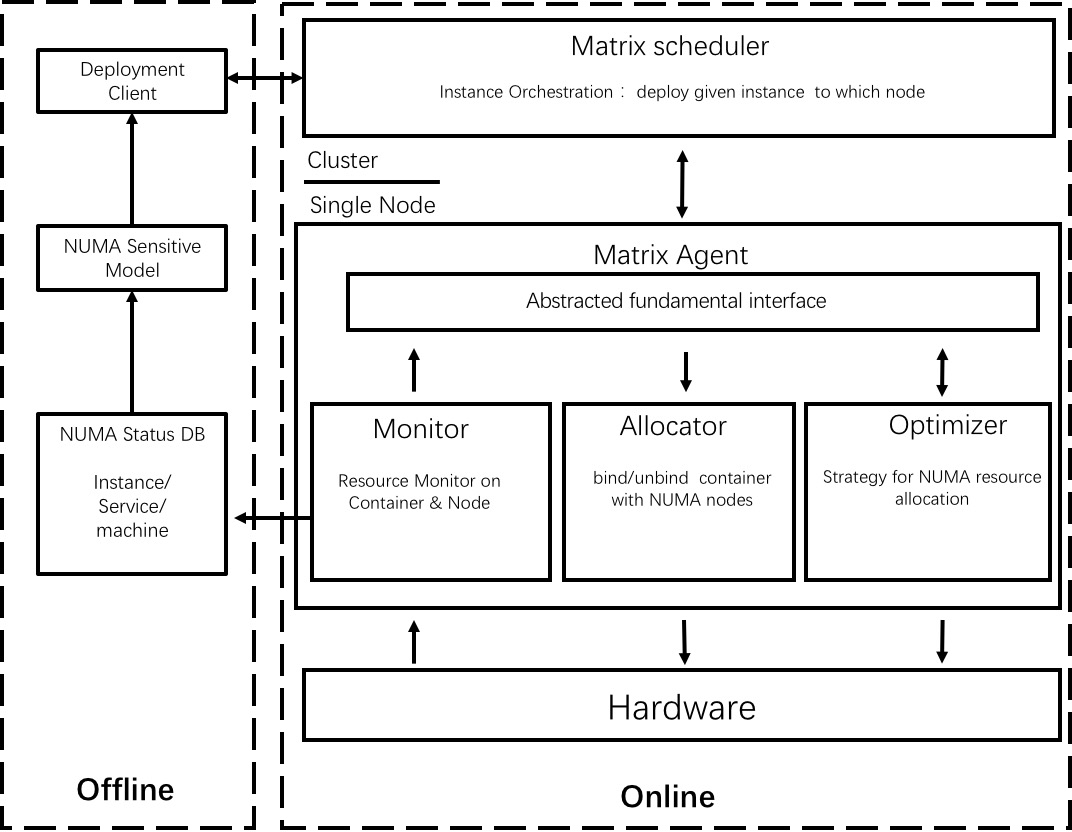}
    \caption{MAO (Memory Access Optimization) platform conceptual architecture diagram}
    \label{fig:numa_node_overview}
\end{figure}

As figure5 indicates, MAO solution comprises two parts: 1) offline storage, analysis for workloads’ NUMA sensitivity character and 2) online optimization which invokes functions from single node agent/actuator and cluster scheduling. 
Within online optimization part, there are different modules like Matrix agent, resource monitor, allocator and resource optimizer. Among those modules:
\begin{itemize}
  \item \textbf{Matrix Agent} acts as the interface to conduct node level optimization. It will get numactl parameters based on hints from Cluster scheduler and offline Analyzer to conduct node level resource allocation. 
  \item \textbf{Monitor} will be responsible for collecting the resource consumption statistic of the node and containers that run within that. The data will be used by Matrix cluster orchestration platform, offline Analyzer and Resource Optimizer.  
  \item \textbf{Allocator} serves for conducting NUMA bind/unbind operation of the workload container module to the NUMA nodes based on the request.   
  \item \textbf{Optimizer} will conduct NUMA runtime optimization. It will check whether the memory access local ratio is optimal to reach the target and trigger NUMA bind/unbind operation based on hardware performance monitoring data collected from Monitor and parameters provided by Offline module. In future, if there’s still optimization opportunities like resource is saturated, other optimization actions like node resource throttling or memory page duplication/migration will be conducted.   
\end{itemize}
The offline workload characterization module in MAO will proceed on the data analysis and resource sensitivity model training. The modules include:
\begin{itemize}
\item \textbf{NUMA status database} stores the metrics from \textbf{Online Monitor} to different node/instance and aggregated by workload service level granularity. The database can also support different queries to end users.
\item \textbf{NUMA sensitivity module} implements a ML based feature to characterize each workload service and provide the benefit prediction based on the each service.
\item \textbf{Deployment client} select top workload services based on NUMA sensitivity model result and conduct NUMA bind/unbind optimization. The client also support additional functions like user notification, operation rollback, etc.

\end{itemize}

\subsection{Online: Matrix scheduler and Matrix Agent}
Matrix as the cornerstone to Baidu's on premises cluster management system, supports large-scale application mix-deployment across the data center. Similar systems are currently in addition to Google Borg/Omega \cite{verma2015large}. Currently, the Matrix has hosted all of Baidu's online/offline computing (including Agent computing) and distributed storage, as well as feed, search, e-commercial systems, and most of the Baidu’s core infrastructure systems. 
There are Matrix scheduler and Matrix Agent supporting the cluster orchestration. Matrix scheduler will get NUMA modeling parameters from offline characterization and conduct NUMA optimization hints based on that. Matrix Agent will be deployed into all individual servers and act as the interface to communicate with Matrix scheduler in cluster level. 

\subsection{Online: Monitor}
Regarding the data feed into offline Analyzer, we leverage the Monitor module in each server to continuously collect workload related metrics by using hardware performance counters\cite{psaroudakis2016adaptive}, the key metrics definition with collected target and which tools to get described as below:
\begin{table}[h!]
\centering
\resizebox{8cm}{1.2cm}{
\begin{tabular}{||c | c | c ||} 
 \hline
 Metrics & Target & Tools integrated\\ [0.25ex] 
 \hline\hline
 Memory Bandwidth   & Container & Intel rdt+resctrl+Cgroup   \\ 
 Memory Page Layout & Process   & {Linux /proc/[pid]/numa\_maps} \\
 (RSS page ratio)   &  &    \\
 Memory stall cycles& Container & Intel rdt+resctrl + Cgroup \\
 stall cycles total & Container & Intel rdt+resctrl + Cgroup  \\
 Cpu/memory utilization      & Container                 & Intel rdt+resctrl + Cgroup  \\
 \hline
\end{tabular}
}
\caption{Metrics collected by Monitor for NUMA sensitivity model}
\label{table:data}
\end{table}

\subsection{Online: Allocator}
Allocator is the module to conduct the NUMA binding actions for different workloads. \\
\textbf{Why need additional NUMA bind:} \\
When NUMA configuration is enabled, the memory allocation strategy will use “first touch” policy by default. Under this policy, the process that first touches (that is, writes to, or reads from) a page of memory causes that page to be allocated in the node on which the process is running. This policy works well for some sequential programs and for many parallel programs as well. But performance drop is observed in some other scenarios. For example, due to kernel scheduler’s load balance strategy, the workload process might be migrated to other NUMA nodes for the system level computing efficiency. This will make the local memory access become remote memory access. Another case could be in multi-thread applications, since it is possible that memory allocation is across the NUMA nodes due to different threads allocated in different NUMA nodes. Such scenarios will make the default “first touch” policy not efficient enough. \\
\textbf{What is runtime bind/unbind strategy:} \\
NUMA node binding refers that a workload is leveraging syscall or cgroup/cpuset subsystem to implement a policy run within identified core range and memory region that associated. In Baidu, we use cgroup/cpuset to implement the NUMA binding/unbinding.
After the instance is scheduled to a specific server, Matrix scheduler will make the decision to identify specific NUMA node for that instance based on the statistics of the left resources, including CPU quota, memory size, memory bandwidth and disk capacity, etc. 
During the runtime life cycle of the workload container, the unbind/rebind operation will follow the decision get from optimizer in real time.

\subsection{Online: Optimizer}
Optimizer module will be responsible for the runtime optimization, target for container and resource utilization optimization within the same server node to make sure the best result is sustainable for MAO solution. 

\textbf{NUMA unbind/re-bind optimization for unbalance scenario:}
Sometimes, due to the static NUMA node binding, The load of  the NUMA nodes on a server may become unbalanced and this will make the additional complexity to the Matrix orchestration scheduler. To mitigate the frequent application migration and rescheduling, Optimizer will select instance with least CPU consumption and conduct NUMA unbind operation until load is balanced within the system. In this case, we developed two different strategy as below. In production, When we change the strategy from A to B, the ratio of unbind drops significantly.
\begin{itemize}
    \item A: trigger unbind only if all conditions met: (1) When the CPU quota of the container is over utilized, aka CPU used percent.metrics \textgreater 100\%.  (2) When the container is binded to the NUMA node and CPU utilization of that NUMA node \textgreater 80\%.
    \item B: trigger unbind only if all conditions met: (1) unbalance: when CPU utilization in one NUMA Node is much higher than other NUMA node (The threshold based on historical CPU utilization statistic, we take 10-15\% here). (2) NUMA node hotspot:When the container is binded to the Node and CPU utilization of that node \textgreater P95 historical CPU utilization statistic, (3)  When the CPU quota of the instance is over utilized, aka CPU used percent.metrics \textgreater 100\%
\end{itemize}
\begin{figure}[h]
    \centering
    \includegraphics[width=0.45\textwidth]{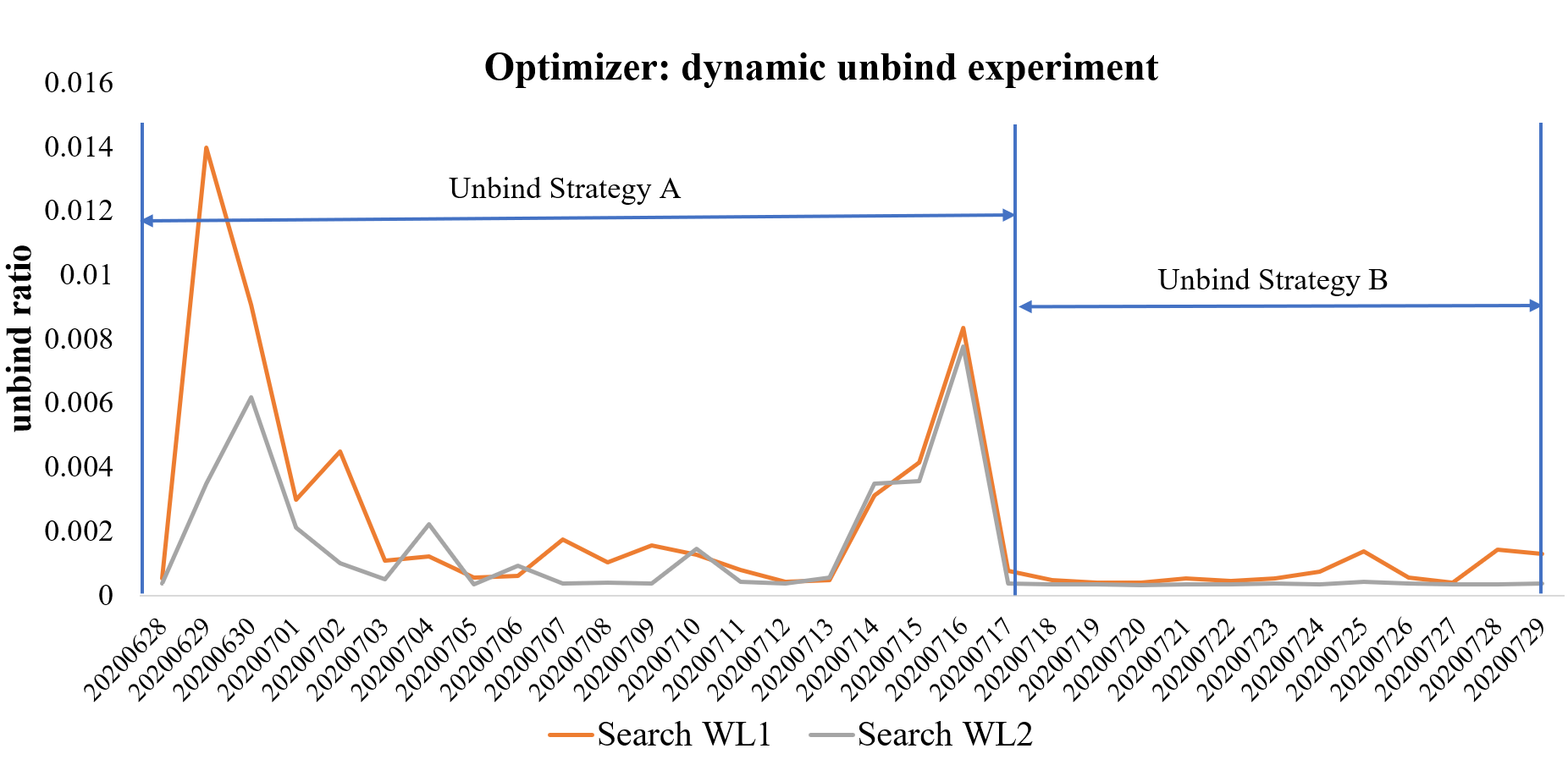}
    \caption{The trend with different condition to conduct unbind optimization}
    \label{fig:unbind optimization}
\end{figure}
 Figure 6 illustrated the experiment for strategy A and B in production for two search workload. We defined unbind ratio as unbinded instance times divide the overall run time. As you see in the chart, we applied strategy A before July 19, 2020 and moved strategy B after that. Obviously strategy B is better and finally used in production because of less unbind in production while it very well supported search SLA (service level agreement).
 
The optimizer process is shown in Figure 7. Firstly it will obtain the instance and server information, and then determine whether unbind is required according to the above unbind conditions, if not, determine whether there is instances that need to be bind for NUMA optimization. Finally it sends commands to the Allocator for execution.
\begin{figure}[h]
    \centering
    \includegraphics[width=0.45\textwidth]{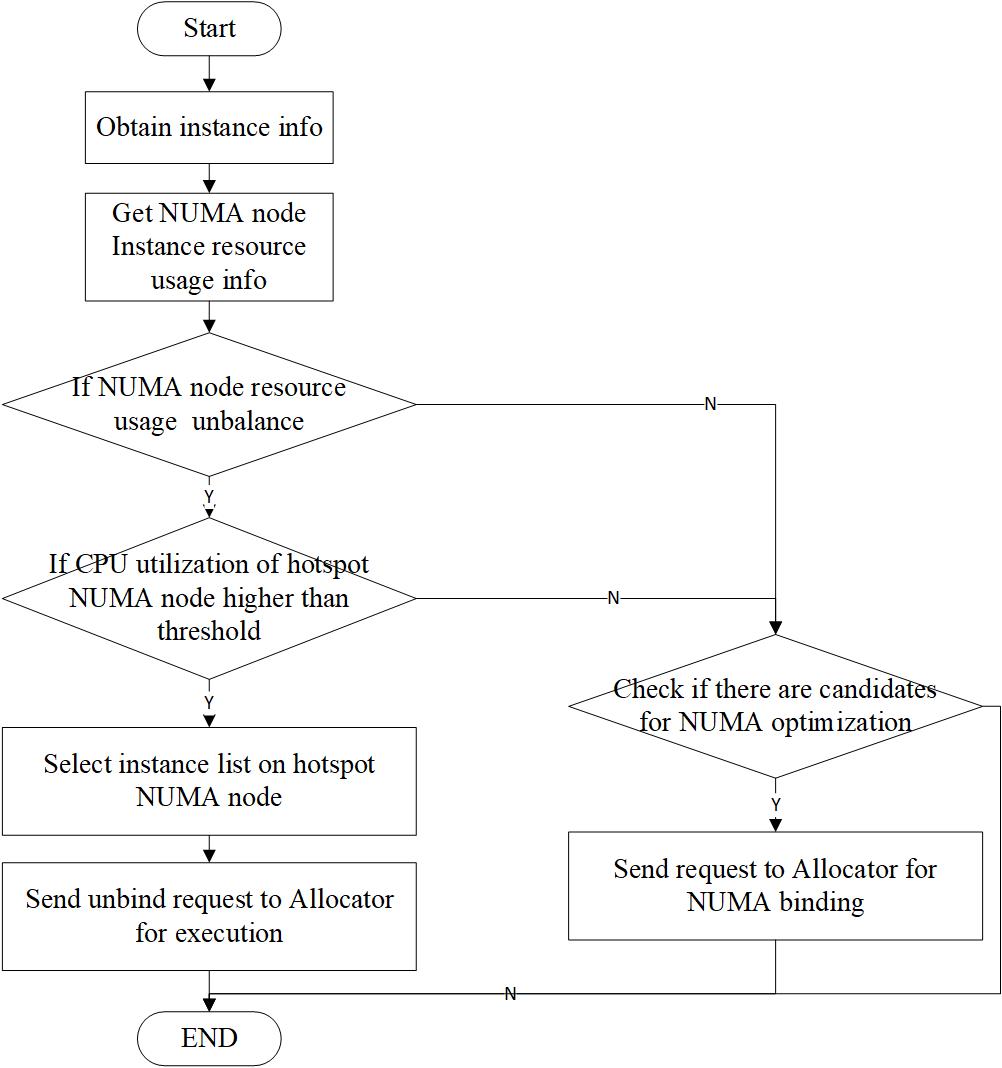}
    \caption{Optimizer workflow to dynamic adjust the resource policy}
    \label{fig:Optimizer workflow}
\end{figure}


\subsection{Offline: NUMA sensitivity Modeling}

\subsubsection{NUMA sensitivity model definition}
Whether a workload can benefit from NUMA or not will base on the workload performance relevance to the memory locality. We summarized 4 performance metrics to build up the performance model for workloads as the criteria to conduct NUMA binding for those workloads in Table 4. 

\begin{table}[h!]
\centering
\resizebox{8cm}{1.2cm}{
\begin{tabular}{||c | c ||} 
 \hline
 NUMA sensitivity model & Details \\
 \hline\hline
 MBW    & Memory (DRAM) Bandwidth  \\ 
 MSR    & Memory (DRAM) Stall cycles Ratio  \\
 NPMR   & Non Page-cache Memory Ratio   \\
 RMBR   & Remote Memory Bandwidth Ratio \\
 \hline
\end{tabular}
}
\caption{NUMA sensitivity model based on MBW, MSR, NPMR and RMBR}
\label{table:data}
\end{table}

\begin{itemize}
    \item \textbf{Memory bandwidth - MBW}
    The workload’s memory bandwidth consumption indicates the memory access amount within a certain period of time. It also reflects the dependency on the memory resource of the workloads. Within the same local/remote ratio of memory access, the more consumption of memory bandwidth, the more remote access amount the workload may have to suffer, which indicates the more benefits with NUMA binding. Also, if the workload’s performance is more sensitive to the memory resource, the benefit of the NUMA optimization will be more significant and this metric is considered as a positive factor.
    \item \textbf{Memory (DRAM) Stall cycles Ratio - MSR}
    \[ MSR =  \frac{Memory\_Stall\_Cycles}{CPU\_CLK\_UNHALTED.THREAD} \]
    
    \textit{Memory stall cycles} indicates the cycles that workloads need to wait for the memory Read/Write operation to complete. So the metric reflects the possibility of system memory becomes the bottleneck of the workload performance. The larger value means the workload is stalled more on memory operation, which brings more performance impact to the workload. Compared with \textit{memory bandwidth (MBW)} metric, this metric consider about both the access latency and access amount which would impact the workload latency, meanwhile the \textit{memory bandwidth (MBW)} is major focus on the QPS of the workload. So metric \textit{memory stall cycles/stall cycles total} is also a positive factor.
    \item \textbf{Non page-cached memory allocation ratio}
    \[ NPMR =  \frac{NUMA.Page\_RSS}{NUMA.Page\_Total} \]
    
    The type of memory pages workloads involve will impact locality of the workload memory access. Memory pages will be categorized as mmapped pages, aka RSS(resident set size) which include anonymous page(file map) and unmapped cache pages (page cache). Mapped pages has virtual address, meanwhile page cache belongs to unmapped page which does not have virtual address and allocation is decided by operate system instead of workload itself. During the workload running, when more and more page cache is allocated, less memory requests of the workload would be allocated locally. \par
    The metric indicates the ratio of the workload memory allocation (non page-cached) to the total memory allocation. Since the memory pages which belongs to anonymous page will not be optimized from NUMA binding, the solution needs to consider about this to be part of the factor. So the larger value the metric reports, the more headroom available for NUMA optimization. For the mapped pages, we will consider to leverage hot page migration capability from Online Optimizer in the future. \par
    \item \textbf{Remote memory bandwidth ratio - RMBR}
     \[ RMBR =  \frac{Remote\_Memory\_bandwidth}{Total\_Memory\_Bandwidth} \]
    
    This metric indicates the remote memory bandwidth ratio out of the total memory bandwidth for a specific workload. The more remote memory bandwidth, the more benefits we may get from NUMA optimization.
\end{itemize}

\subsubsection{NUMA sensitivity model Implementation} \par
The task of building performance models for predicting the impact of NUMA optimization falls into the category of supervised learning. In order to capture training dataset, we conducted NUMA optimization on selected containers for various applications in production and collected performance metrics with their performance improvement. 
We leveraged XGBoost library\cite{chen2016xgboost} for the model training. XGBoost is an optimized distributed gradient boosting library designed to be highly efficient, flexible and portable. It implements machine learning algorithms under the Gradient Boosting framework and has been proved in practice that it can be effectively used for classification and regression tasks.\cite{nielsen2016tree} Tree boosting methods have empirically proven to be a highly effective and versatile approach to predictive modeling. XGBoost has gained popularity by winning numerous machine learning competitions, and becomes a highly adaptive method. 
In this case, the objective is to predict possible improvement with NUMA optimization via regression based on the 4 performance metrics we discussed. We generate the model based on XGBoost and use GridSearchCV from SKLearn library \cite{louppe2016introduction} to fine tune those hyper parameters. We finally select the model that best fits our data using the mean absolute error (MAE) and the coefficient of determination R2. 


\section{RESULTS}

\subsection{Negligible online adoption overhead}
The overhead evaluation is a key indicator for the implemented solution to be deployed into an online production environment. During the MAO implementation, we also carefully evaluate the overhead of each online module to make sure it won’t bring any performance impact to the business workloads. As a result, we summarized the overhead conclusion for each sub module as below:
\begin{itemize}
    \item Matrix scheduler and agent: the scheduler and agent belongs to infrastructure framework, the additional API or communication overhead is negligible.
    \item Monitor: PMU based monitoring will takes around 0.06\% of a single hyperthread of the CPU. Meanwhile NUMA sensitivity related metrics monitoring will take about 0.48\% of a single hyperthread of the CPU. The added up CPU overhead is around 0.5\% of a single hyperthread which is acceptable to an online monitoring module from our perspective.
    \item Allocator: Identified the NUMA topology or each machine and conduct cgroup/cpuset with the NUMA node and core core. Since this is not a periodically task, the overhead is also negligible.
    \item Optimizer: Periodically conduct module dynamic NUMA bind operation based on the modules list identified to each machine. The overall execution each round will be completed within less than 1ms and average CPU utilization cost is 9.4\% of a single hyper thread of the CPU.
\end{itemize}
After all, since MAO can be deployed in non-intrusive manner, no need server reboot or workload service restart. Given the overhead brought by MAO is very little, it is pretty acceptable to be deployed online to the production environment. 

\subsection{Widely deployed to Search and Feed clusters with significant performance gain}
 
Search and Feed are key services in Baidu’s product line, which also serves many other products like Mobile, Poster, Map, etc. With NUMA sensitivity model defined, we list out the major modules in Feed and Search workloads. In order to quantify the benefits, we select top 4 functional modules cover embedding to influence called Feed-Predictor1, Feed-Predictor2, Router and Feature Service. These services has deployed across 4 data centers with totally 12.7K servers. As Figure8 shows, the result in Feed and Search workloads looks promising.
\begin{figure}[h]
    \centering
    \includegraphics[width=0.45\textwidth]{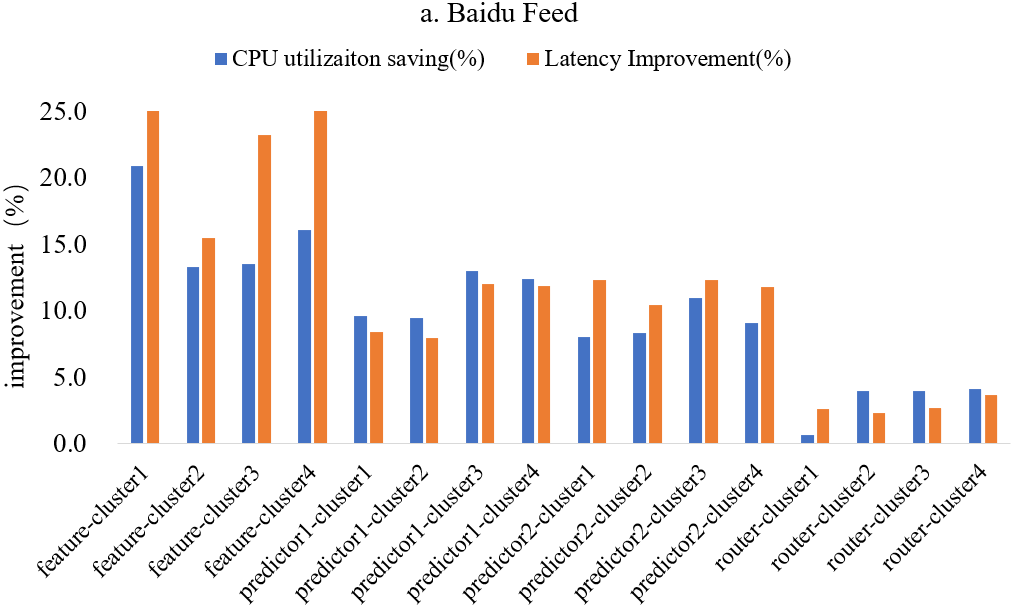}
    \includegraphics[width=0.45\textwidth]{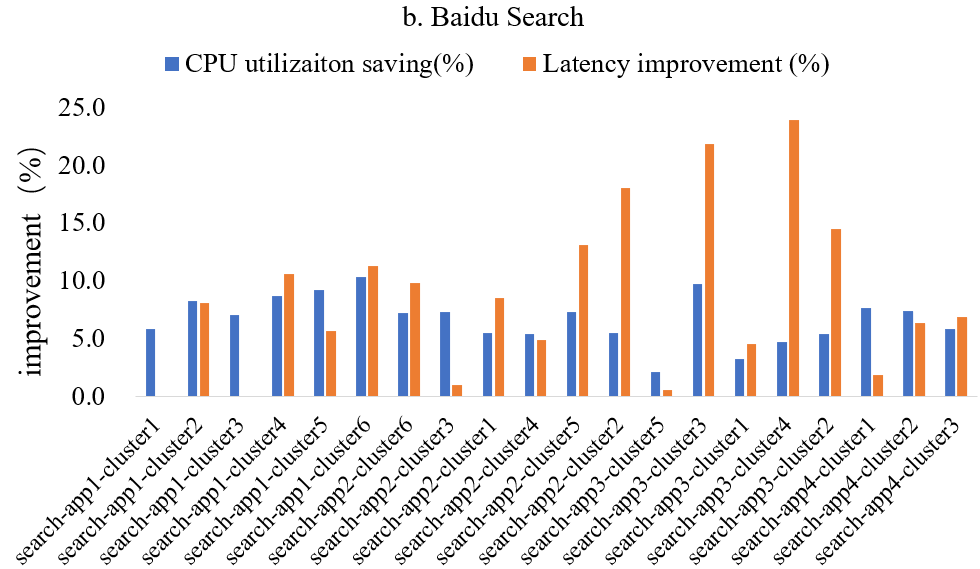}
    \caption{Feed and Search workload optimization benefits from MAO}
    \label{fig:different NUMA sensitive module}
\end{figure}
    With MAO, we have achieved 10.1\% for Predictor1, 11.72\% for Predictor2, 2.8\% for Router and 23.8\% for Feature Service on average latency improvement accordingly which leads to overall 12.1\% average latency improvements and 9.8\% CPU resource saving. While for Search, we have similar 8.6\% improvements in average latency and 6.7\% CPU resource saving respectively. \\

    We further compared basic micro architecture performance metrics like memory locality and IPC for Search workload by comparing with the case without NUMA aware optimization (but AutoNUMA balancing). Figure9 shows CDF (Cumulative Distribution Function) chart for container count with NUMA locality improvement in percentage. The light green bar is the container numbers for specified NUMA locality improvement (in X-axis) and the red line is the CDF of improvement. As a result we observed 17.26\% average improvements in local memory access ratio and 8.8\% improvements in IPC respectively.
    
\begin{figure}[h]
    \centering
    \includegraphics[width=0.45\textwidth]{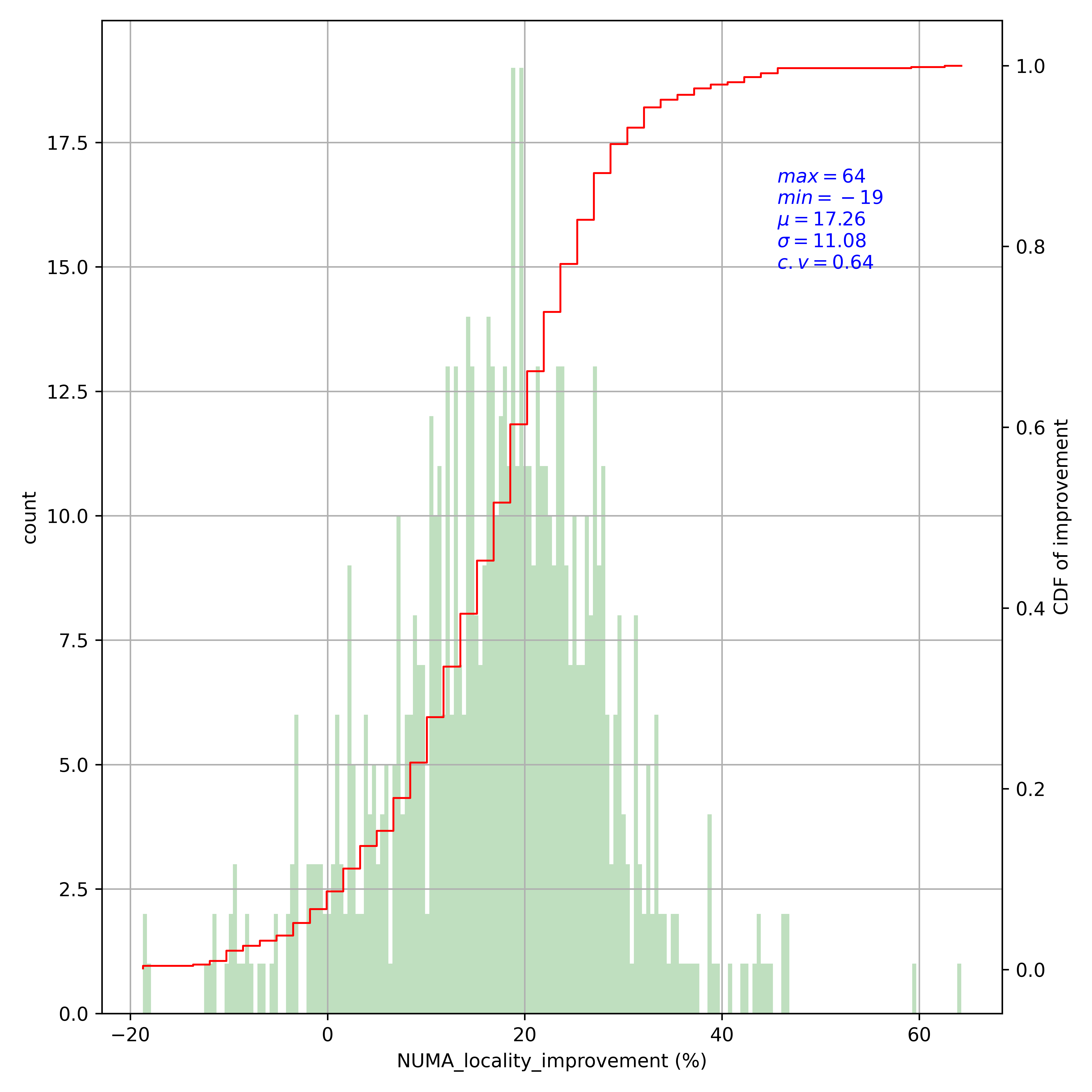}
    \caption{Search workload optimization benefits from MAO}
    \label{fig:Search workload optimization benifits from MAO}
\end{figure}

\subsection{Good predication accuracy with NUMA sensitivity model}
With the discussion on Section 3.5 for the methodology to build up NUMA sensitivity model, We collected 2000 container based samples with NUMA optimization in production from different cloud applications (different services in Search and Feed) and further randomly split 85\% of samples for training and 15\% for prediction (also known as test samples). In order to compare the model performance on other well know regression libraries, we trained 3 different models here, including RandomForestRegressor and LinearRegression as well. 
Finally we got 0.92, 0.87 and 0.19 on R2 on MAE respectively for XGBoost, RandomForestRegressor and LinearRegression library that shows XGBoost performs the best among all these libraries, while LinearRegression failed to fit well because obviously it is not a linear problem. 
Figure 10 shows the feature importance for Case 1 and Case 2, among all the 4 features (performance metrics), we could see MBW and RMBR is a little more important than the other two. The more RMBR and MBW, the higher NUMA sensitivity is expected, but overall these 4 metrics together contributed the NUMA sensitivity. 

\begin{figure}[h]
    \centering
    \includegraphics[width=0.45\textwidth]{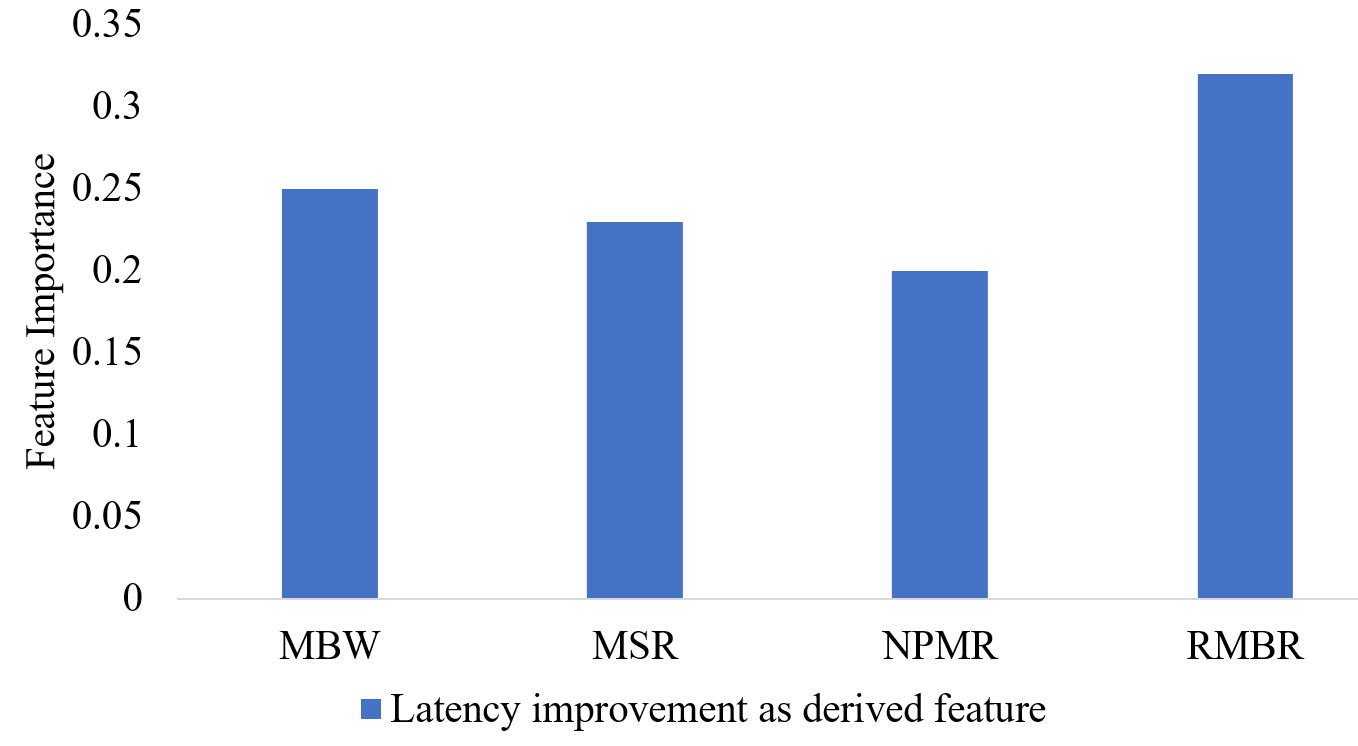}
    \caption{Feature importance of Case1 and Case2}
    \label{fig:feature Importance}
\end{figure}

\begin{figure}[h]
    \centering
    \includegraphics[width=0.45\textwidth]{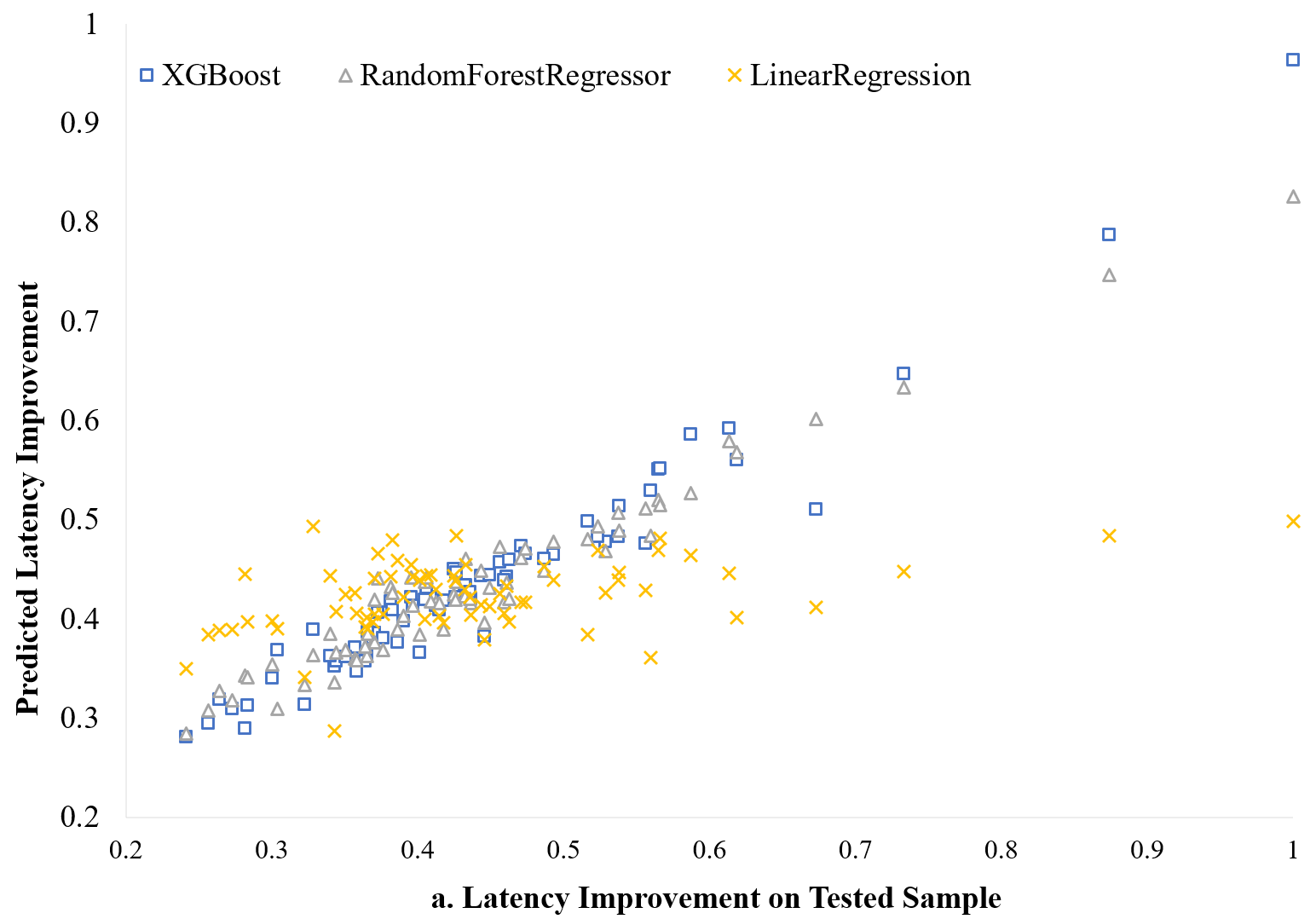}
    \includegraphics[width=0.45\textwidth]{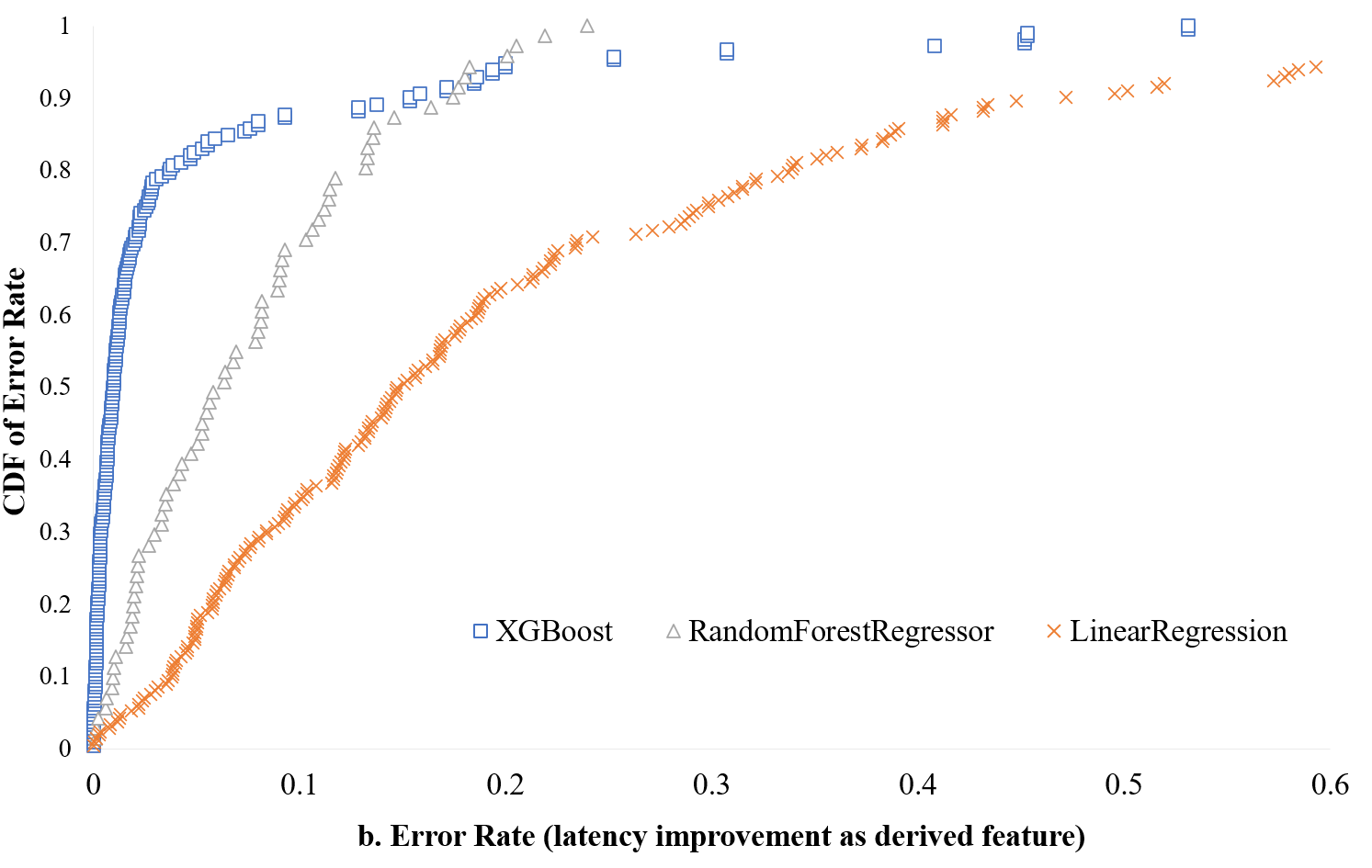}
    \caption{Accuracy of NUMA sensitivity model (Search workload)}
    \label{fig:Accuracy of NUMA sensitive model (Feed workload)}
\end{figure}

Figure 11 is the CDF distribution for different Error Rate and the real values from predicted latency improvement via NUMA optimization compared with real improvement collected in production. In all of 300 test samples, we only see 0.025 MAE error rate for XGBoost model which is better than other two models of RandomForestRegressor and LinearRegression, at 0.035 and 0.086 respectively, the LinearRegression performance are the worse since such kind of problem is not a linear problem. As a result, we conclude our NUMA sensitivity model can be used accurately to predict performance improvement for NUMA optimization.

\section{DISCUSSION}
Beyond current implementation of MAO, there are some other optimization options we are considering for the future online optimization to adopt. \\  
\textbf{Contention Isolation:} \\
Even with the optimization from MAO, we do observed more workload modules be scheduled and binded to the same NUMA node with shared resource running. This will cause more resource contention than in the production environment when MAO is not deployed. The next step of MAO to protect the resource of each workload module is called Contention Isolation. The logical design is when we found performance of the workload drops significantly but CPU utilization is not that high, we may check if there’s any contention happens within the micro architecture resource like memory bandwidth and last level cache contention. If there’s memory bandwidth contention or cache trash cases happen, we would leverage Intel CAT(Cache Allocation Technology) and (MBA) Memory Bandwidth Allocation, aka Intel Resource Direct Technology feature set \cite{intel-RDT} as a good complementary to protect these resource consumption from contention of the other aggressive or low priority workloads. \\
\textbf{Hot memory page migration:} \\
With our current kernel’s implementation, the total memory consumption will become bigger and bigger along with the page cache’s increase. As a result, workload will have to conduct more memory allocation to the remote nodes sooner or later. The access of the workload to remote memory will increase the average memory access latency which eventually impact the performance as well. As the mitigation of this, Optimizer will use hot page migration to conduct run time optimization to workload to have best locality. Different from page migration strategy of AutoNUMA, this migration will focus on workload's working set size related memory hot page identification that is critical to the performance comparing to the migration overhead.  
Eventually, if the Optimizer module found performance is still not back to the optimal after above optimization has been done, it will report out the workload information and results to Matrix scheduler. And then, Matrix scheduler will decide whether to migrate or kill the containers within the server node. The current impediments to be adopted in online production environment is due to lack of enough information to build up a practical migration algorithm to get clear benefits projection comparing to the cost of page migration. Related work introduced in \cite{lepers2014improving} has shared with us some initial ideas on the implementation of next step.

\section{CONCLUSION}
Modern multi-core systems are based on a Non-Uniform Memory Access (NUMA) architecture and NUMA systems are increasingly common. \par
We implement a NUMA optimized holistic solution called MAO for online production deployment. The NUMA awareness is created based on the real problem statement we have observed in the online complex workload running environment, that only NUMA bind or AutoNUMA may not be able to meet our expectations. \par
By collecting the workload characterization data with build up the NUMA sensitivity model, we can identify top workload modules that should be essential to be optimized based on NUMA awareness. With the NUMA allocation and real-time dynamic optimization, we have successfully deployed MAO to over one million servers in Baidu’s online fleet environment. Some good progress has been made, for example, in Feed product, we have achieved 12.1\% average latency improvements and 9.8\% CPU resource saving. The promising result successfully demonstrates the effectiveness of this solution and become worthy to widely scale out to more workloads coverage within Baidu’s WSC.

\printbibliography 

@inproceedings{tang2013optimizing,
  title={Optimizing Google's warehouse scale computers: The NUMA experience},
  author={Tang, Lingjia and Mars, Jason and Zhang, Xiao and Hagmann, Robert and Hundt, Robert and Tune, Eric},
  booktitle={2013 IEEE 19th International Symposium on High Performance Computer Architecture (HPCA)},
  pages={188--197},
  year={2013},
  organization={IEEE}
}

@article{corbet2012autonuma,
  title={AutoNUMA: the other approach to NUMA scheduling},
  author={Corbet, Jonathan},
  journal={LWN. net},
  year={2012}
}

@inproceedings{blagodurov2010case,
  title={A case for NUMA-aware contention management on multicore systems},
  author={Blagodurov, Sergey and Fedorova, Alexandra and Zhuravlev, Sergey and Kamali, Ali},
  booktitle={2010 19th International Conference on Parallel Architectures and Compilation Techniques (PACT)},
  pages={557--558},
  year={2010},
  organization={IEEE}
}

@phdthesis{lepers2014improving,
  title={Improving performance on NUMA systems},
  author={Lepers, Baptiste},
  year={2014}
}

@article{psaroudakis2016adaptive,
  title={Adaptive NUMA-aware data placement and task scheduling for analytical workloads in main-memory column-stores},
  author={Psaroudakis, Iraklis and Scheuer, Tobias and May, Norman and Sellami, Abdelkader and Ailamaki, Anastasia},
  journal={Proceedings of the VLDB Endowment},
  volume={10},
  number={2},
  pages={37--48},
  year={2016},
  publisher={VLDB Endowment}
}

@inproceedings{brecht1993importance,
  title={On the importance of parallel application placement in NUMA multiprocessors},
  author={Brecht, Timothy},
  booktitle={Symposium on Experiences with Distributed and Multiprocessor Systems (SEDMS IV)},
  pages={1--18},
  year={1993}
}

@inproceedings{denoyelle2019data,
  title={Data and thread placement in NUMA architectures: A statistical learning approach},
  author={Denoyelle, Nicolas and Goglin, Brice and Jeannot, Emmanuel and Ropars, Thomas},
  booktitle={Proceedings of the 48th International Conference on Parallel Processing},
  pages={1--10},
  year={2019}
}

@inproceedings{funston2018placement,
  title={Placement of Virtual Containers on $\{$NUMA$\}$ systems: A Practical and Comprehensive Model},
  author={Funston, Justin and Lorrillere, Maxime and Fedorova, Alexandra and Lepers, Baptiste and Vengerov, David and Lozi, Jean-Pierre and Qu{\'e}ma, Vivien},
  booktitle={2018 $\{$USENIX$\}$ Annual Technical Conference ($\{$USENIX$\}$$\{$ATC$\}$ 18)},
  pages={281--294},
  year={2018}
}

@article{guide2011intel,
  title={Intel{\textregistered} 64 and ia-32 architectures software developer’s manual},
  author={Guide, Part},
  journal={Volume 3B: System programming Guide, Part},
  volume={2},
  pages={11},
  year={2011}
}

@inproceedings{verma2015large,
  title={Large-scale cluster management at Google with Borg},
  author={Verma, Abhishek and Pedrosa, Luis and Korupolu, Madhukar and Oppenheimer, David and Tune, Eric and Wilkes, John},
  booktitle={Proceedings of the Tenth European Conference on Computer Systems},
  pages={1--17},
  year={2015}
}

@article{viswanathan2013intel,
  title={Intel memory latency checker},
  author={Viswanathan, Vish and Kumar, Karthik and Willhalm, T and Lu, P and Filipiak, B and Sakthivelu, S},
  journal={Intel Corporation},
  year={2013}
}

@inproceedings{goglin2009enabling,
  title={Enabling high-performance memory migration for multithreaded applications on linux},
  author={Goglin, Brice and Furmento, Nathalie},
  booktitle={2009 IEEE International Symposium on Parallel \& Distributed Processing},
  pages={1--9},
  year={2009},
  organization={ieee}
}

@inproceedings{antony2006exploring,
  title={Exploring thread and memory placement on NUMA architectures: Solaris and Linux, UltraSPARC/FirePlane and Opteron/HyperTransport},
  author={Antony, Joseph and Janes, Pete P and Rendell, Alistair P},
  booktitle={International Conference on High-Performance Computing},
  pages={338--352},
  year={2006},
  organization={Springer}
}

@inproceedings{majo2011memory,
  title={Memory system performance in a NUMA multicore multiprocessor},
  author={Majo, Zoltan and Gross, Thomas R},
  booktitle={Proceedings of the 4th Annual International Conference on Systems and Storage},
  pages={1--10},
  year={2011}
}

@techreport{mccormick2011empirical,
  title={Empirical memory-access cost models in multicore numa architectures},
  author={McCormick, Patrick S and Braithwaite, Ryan Karl and Feng, Wu-chun},
  year={2011},
  institution={Los Alamos National Lab.(LANL), Los Alamos, NM (United States)}
}

@inproceedings{luo2016compositional,
  title={Compositional model of coherence and NUMA effects for optimizing thread and data placement},
  author={Luo, Hao and Brock, Jacob and Li, Pengcheng and Ding, Chen and Ye, Chencheng},
  booktitle={2016 IEEE International Symposium on Performance Analysis of Systems and Software (ISPASS)},
  pages={151--152},
  year={2016},
  organization={IEEE}
}

@inproceedings{arapidis2018performance,
  title={Performance Prediction of NUMA Placement: A Machine-Learning Approach},
  author={Arapidis, Fanourios and Karakostas, Vasileios and Papadopoulou, Nikela and Nikas, Konstantinos and Goumas, Georgios and Koziris, Nectarios},
  booktitle={2018 IEEE International Conference on Cloud Computing Technology and Science (CloudCom)},
  pages={296--301},
  year={2018},
  organization={IEEE}
}

@article{gaud2015challenges,
  title={Challenges of memory management on modern NUMA systems},
  author={Gaud, Fabien and Lepers, Baptiste and Funston, Justin and Dashti, Mohammad and Fedorova, Alexandra and Qu{\'e}ma, Vivien and Lachaize, Renaud and Roth, Mark},
  journal={Communications of the ACM},
  volume={58},
  number={12},
  pages={59--66},
  year={2015},
  publisher={ACM New York, NY, USA}
}

@inproceedings{chen2016xgboost,
  title={Xgboost: A scalable tree boosting system},
  author={Chen, Tianqi and Guestrin, Carlos},
  booktitle={Proceedings of the 22nd acm sigkdd international conference on knowledge discovery and data mining},
  pages={785--794},
  year={2016}
}

@mastersthesis{nielsen2016tree,
  title={Tree boosting with xgboost-why does xgboost win" every" machine learning competition?},
  author={Nielsen, Didrik},
  year={2016},
  school={NTNU}
}

@misc{intel-RDT,
  title = {{Intel Resource Direct Technology }},
  howpublished = {https://www.intel.com/content/www/us/en/architecture-and-technology/resource-director-technology.html}
  }

@article{louppe2016introduction,
  title={An introduction to Machine Learning with Scikit-Learn},
  author={Louppe, Gilles},
  year={2016}
}


\end{document}